\documentclass[
preprint,
 amsmath,amssymb,
aps,
]{revtex4-2}

\usepackage{graphicx}
\usepackage{dcolumn}
\usepackage{bm}
\usepackage{hyperref}

\begin{document}
\preprint{APS/123-QED}

\title{Spatiotemporally Resolved Measurements of CO$_2$ Distribution at the Air-Water Interface Using Tunable Diode Laser Spectroscopy}
\author{Dongfang Zhao, Yumin Shi, Shengkai Wang}
\thanks{Corresponding author: sk.wang@pku.edu.cn (Shengkai Wang)}
\affiliation{SKLTCS, CAPT, School of Mechanics and Engineering Science, Peking University, 5 Yiheyuan Road, Haidian District, 100871, China}%
\date{\today}
             
\begin{abstract}
The transport of CO$_2$ across the air–water interface is central to physical oceanography and carbon sequestration. A comprehensive understanding of this process requires high-resolution diagnostics of diffusion, absorption, and reaction across a wide range of spatial and temporal scales. The current study presents a novel measurement method to quantify the CO$_2$ distribution at the air-water interface. This method combines the advantages of tunable diode laser spectroscopy and rapid spatial beam scanning for in situ, nonintrusive, and spatiotemporally resolved measurement of the CO$_2$ concentration distribution above the interface. The performance of this method was examined in a series of quasi-1D experiments in a miniature gas chamber, where the diffusion and absorption of CO$_2$ into pure water and alkaline solutions of different pH values were continuously monitored. An effective time resolution of 5 ms and an effective spatial resolution of 1 mm were achieved. The observed gas-phase CO$_2$ distribution evolution agreed with the classic one-dimensional diffusion model, which validated the accuracy of the current method. PH-dependent dynamics of interfacial CO$_2$ concentration was also observed: the CO$_2$ depletion rate is highly pH-sensitive at low pH and saturates at pH $\approx$ 10, revealing complex competition between the gas-phase and the liquid-phase transport processes. The current method's high spatial and temporal resolution holds promise for studying cross-interface gas transport under more complex flow conditions, in both field measurements and laboratory studies.
\end{abstract}

\maketitle
\clearpage

\section{Introduction}
The exchange of CO$_2$ across the air-water interface has attracted intensive scientific interest for more than half a century \cite{liss1974flux, watson1991air, wallace1992large,  schmittner2013biology, yang2021natural}, particularly because of its importance in the mitigation of the anthropogenic greenhouse effect through the oceans' absorption of excess atmospheric CO$_2$ \cite{sarmiento1996oceanic, wanninkhof2009advances}. Previous studies on the interfacial transfer processes of small gas molecules, such as CO$_2$, CH$_4$ and O$_2$, have revealed that their exchange rates at the air-water interface are governed by a multitude of physical and chemical processes that are influenced by wind speed, wave dynamics, surfactants, turbulence, temperature, and salinity \cite{liss1986air, frew1997role, olsen2005effect, woolf2016calculation, zhao2018relationship}. These processes exhibit strong spatiotemporal coupling and pronounced nonlinear characteristics, and render their underlying mechanisms are highly complex.

One of the key challenges in understanding these processes is their multiscale nature. Away from the interface, turbulent transfer is typically orders of magnitude greater than molecular transfer, whereas near the interface, molecular transport dominates. This gives rise to viscous and mass boundary layers on both sides of the air–water interface. Depending on the molecular diffusivity and solubility, the thickness of these layers can range from hundreds of micrometers to tens of meters \cite{jahne1998air}. For example, in the transfer of sparingly soluble gases such as CO$_2$ into pure water, most of the resistance occurs within an approximately 200-$\mu m$-thick diffusion sublayer on the water side \cite{walker2008measurement}, and the effective transfer rate across this layer is largely affected by temperature (which governs the molecular diffusivity \cite{groger2011note}) and pH (which modifies the effective solubility, as discussed in \cite{kuss2004chemical}). Furthermore, recent studies have suggested that capillary-gravity waves \cite{jahne1987parameters, saylor1997gas, adler2022laboratory} and/or sub-surface turbulence \cite{jirka2008experiments, bullee2024influence, li2025sub} can significantly enhance the transfer rate by either distorting the diffusion layer or transporting unsaturated fluid parcels to the surface, thereby increasing the effective concentration gradient. For accurate parameterization of the CO$_2$ transfer rate/flux across the air-water interface, high-resolution measurements are critically needed to elucidate the interfacial transport mechanisms \cite{chevallier2014toward}, as they can lay the foundation for proper up-scaling in micrometeorological models \cite{Fairall2000}.

Representative methods for field measurements of air–water CO$_2$ transport include ship-based eddy covariance \cite{mcgillis2001direct, miller2010ship, dong2021uncertainties}, floating chamber techniques \cite{schrier2010comparison, martinsen2018simple, erkkila2018comparison}, and satellite remote sensing\cite{parard2017remote, yu2023satellite, yuan2025gloflux}. These methods have been successfully employed in relatively large-scale measurements (minutes to hours in time, and tens of meters to kilometers in space) and have provided useful constraints on global carbon budgets. However, they lack the spatial and temporal resolution needed for unraveling the microscale dynamics at the air-water interface, where the dominant resistance and gradient occur. For fundamental studies of gas transfer mechanisms, e.g. in a laboratory, methods of higher resolution and accuracy are needed.

Recent advances in microscale colorimetric and fluorescence imaging have enabled high-resolution visualization of transport processes in the aqueous phase. For example, Cheng and co-workers ~\cite{Cheng2025Taylor} employed micro-PTV, high-speed imaging and a pH-sensitive colorimetric approach to simultaneously measure the liquid-phase velocity field and CO$_2$ concentration distribution within Taylor-flow liquid slugs and investigate the interfacial CO$_2$ transfer behaviors. In another study, Faasen and co-workers \cite{Faasen2024Confinement} used a pH-sensitive fluorophore to visualize the spatiotemporal evolution of CO$_2$ mass boundary layers in a vertical water column and determined the boundary-layer thickness during purely diffusive and convection-dominated stages. Despite the success of these methods in monitoring CO$_2$ transport in the liquid phase, high-resolution diagnostic methods for gas-phase measurements are still scarce. 

Precision measurement capabilities for gas-phase transport processes are critically important -- not only because they provide complementary data for understanding the overall physical picture of cross-interface gas transfer, but also for improved quantification of mass flux. Since the concentration gradient on the gas side is typically orders of magnitude smaller than that on the liquid side, direct determination of flux from gas-phase CO$_2$ distribution measurements in the diffusion sublayer becomes attainable. These capabilities can also advance contemporary efforts in CO$_2$ capture and storage \cite{wang2011post} and sequestration \cite{huppert2014fluid}.

The current study aims to develop a novel diagnostic method for quantifying the CO$_2$ concentration distribution near the air-water interface. This method employs tunable diode laser spectroscopy and rapid spatial beam scanning for high-accuracy CO$_2$ profiling above the interface at millimeter and millisecond resolutions. The remainder of this paper is organized as follows. Section 2 elaborates on the current measurement principle, as well as the technical details of experimental setup and data analysis. Section 3 presents a series of quasi-1D validation experiments in a custom-built CO$_2$ cross-interface diffusion chamber and analyzes the key performance metrics of this method. A parametric study of CO$_2$ transport dynamics at varying pH values is also conducted, and the results are discussed in Section 4. Finally, the concluding remarks and outlook for future work are presented in Section 5.

\section{Methods}
\subsection{Experimental Setup}

\begin{figure}[ht!]
    \centering
    \includegraphics[width = 1 \linewidth]{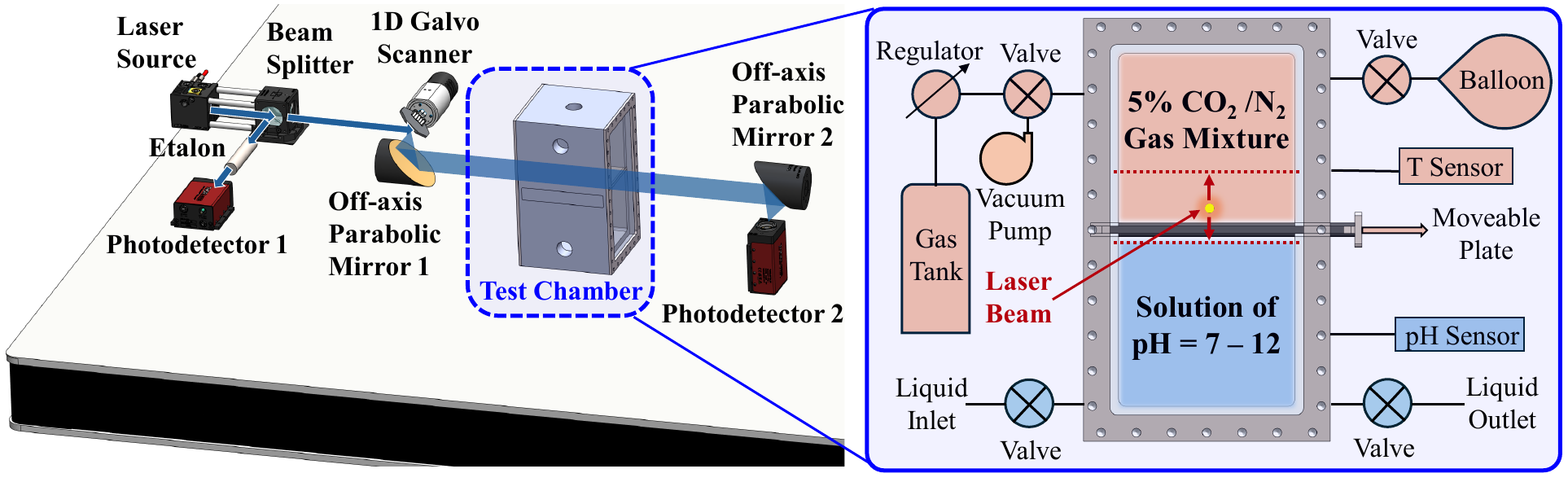}
    \caption{\footnotesize A schematic of the current experimental setup.}
    \label{Fig_01}
\end{figure}

A schematic of the current experimental setup is shown in Fig. \ref{Fig_01}. The experiment was conducted in a custom-built miniature chamber with internal dimensions of 50 mm x 50 mm x 100 mm, with optical access on the front and back sides. A pair of 5-mm-thick calcium fluoride windows was glued to the chamber with epoxy for airtight sealing and compressed by external aluminum frames to enhance mechanical stability. These windows provided more than 90\% transmittance over 200 - 8000 nm, enabling optical measurements across a wide spectral range. A gas mixture of 5\% CO$_2$ in N$_2$ was supplied by a high-pressure cylinder to the test chamber through an inlet valve. The chamber outlet was connected to a balloon reservoir filled with the same gas mixture prior to the experiment, maintaining a constant pressure while CO$_2$ was absorbed into the water. The chamber was also equipped with a thermocouple and a pH probe.

In the middle of the chamber, a thin movable plate was horizontally mounted inside a sliding slot. The plate was press-sealed with a rubber o-ring on its circumference to separate the gas and liquid before the beginning of each experiment. The top chamber was filled with CO$_2$ gas mixture, while the bottom chamber was filled with pure water or NaOH solutions with pH values between 7.5 and 12.0. To start the experiment, the plate was quickly pulled aside, enabling direct contact of the gas with the liquid and initiating cross-interface transport. Throughout the course of the experiment, the gas-phase CO$_2$ concentration distribution was continuously monitored using laser absorption diagnostic, as discussed in the next section. After each experiment, the two halves of the chamber were separated again by the moving plate; the remaining gas in the top chamber was routinely evacuated using a vacuum pump, and the liquid in the bottom chamber was replaced with a fresh solution.

\subsection{Laser Diagnostic}

\begin{figure}[h!]
    \centering
    \includegraphics[width = 0.5\linewidth]{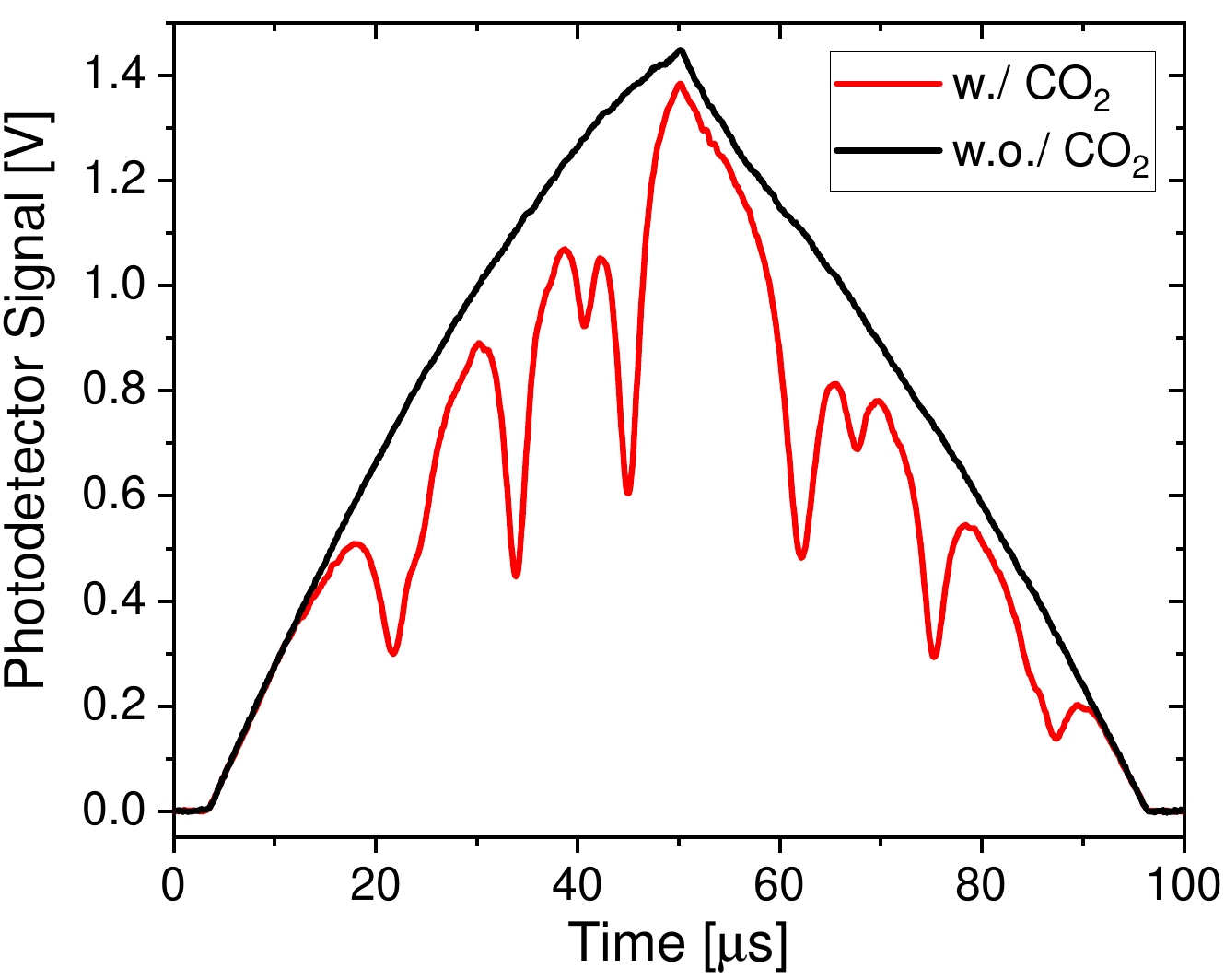}
    \caption{\footnotesize Typical signal obtained in a single laser modulation cycle of 100 $\mu s$.}
    \label{Fig_02}
\end{figure}

The current study employed a narrow-linewidth tunable diode laser absorption diagnostic for quantitative and in situ measurement of gas-phase CO$_2$ concentration. This diagnostic targeted three strong transition clusters near 4360 nm (equivalently, 2294 cm$^{-1}$) in the $\nu_3$ fundamental band of CO$_2$, where interference absorption by water vapor was negligible. These transitions were accessed using a Nanoplus$^{\rm{TM}}$ distributed feedback interband cascade laser (DFB-ICL). The laser output beam was collimated using 1-inch cage mount optics and aligned collinearly with a 650-nm diode laser for visual aid. The beam waist of the ICL was positioned at the center of the test chamber. The $1\sigma$ diameter of the beam waist was measured to be 1.0 $\pm$ 0.1 mm using a translational razor blade, and the Rayleigh range was approximately 0.6 m. To scan over multiple CO$_2$ transitions, the laser was current-modulated using a triangular waveform at 10 kHz by a ppqSense$^{\rm{TM}}$ QubeDL02-T laser controller that was synchronized to a RIGOL$^{\rm{TM}}$ DG2052 function generator (see Fig. \ref{Fig_02} for representative data obtained in a typical laser modulation cycle). The relative change in wavenumber during a current modulation cycle was characterized by a 3-inch germanium etalon with a free spectral range (FSR) of 0.0164 cm$^{-1}$. 

A high-speed parallel beam scanning system was used for spatially resolved measurements. This beam scanning system consisted of a Thorlabs$^{\rm{TM}}$ GVS101 galvo scanner and a pair of off-axis parabolic mirrors (focal length 6 inches, projected diameter 2 inches) placed upstream and downstream of the measurement region. The galvo scanner was electrically modulated by another RIGOL$^{\rm{TM}}$ DG2052 function generator at 100 Hz, rapidly sweeping the reflected beam angle by approximately 7.5 degrees in the vertical direction during each cycle. The output mirror of the galvo scanner was placed right at the focus of the upstream off-axis parabolic mirror; therefore, the reflected beams off the parabolic mirror remained parallel to the horizontal axis. This created a vertical scanning range of the laser beam from -0.5 mm to +5.5 mm relative to the gas-liquid interface. The second off-axis parabolic mirror focused the parallel beams onto a VIGO$^{\rm{TM}}$ PVI-4TE-5 MCT detector. The average transmitted laser power was approximately 3 mW, which was above the saturation intensity of the detector; therefore, a neutral density filter was added in front of the detector to avoid saturation. To eliminate interference from ambient CO$_2$, the entire beam path was enclosed in a plastic container that was continuously purged with nitrogen throughout the experiment.

The transmitted laser intensity and the modulation waveform of the galvo scanner were digitally recorded using a PicoScope$^{\rm{TM}}$ 4424A high-speed data acquisition module at a rate of 10 MS/s. Since each location in the measurement region was traversed twice per scan cycle, an effective full-field measurement rate of 200 Hz (and a single-point measurement rate of 20 kHz) was routinely achieved, corresponding to an effective time resolution of 5 ms. As for the spatial resolution, the nominal precision of the laser beam center location during the vertical scan was approximately 0.1 mm, which is much smaller than the effective beam radius (approximately 0.4 mm HWHM). Formally, the overall spatial resolution is limited by the latter, since high-frequency fluctuations below this length scale would be averaged out. However, under conditions where the spatial distribution of CO$_2$ concentration is monotonic (as in the current study), the actual spatial profiles can be determined from the nominal concentration (obtained from spatially averaged measurement with a Gaussian beam) via deconvolution. In fact, the nominal concentration is usually very close to the actual local value at the beam center, except for some rare cases where an extreme gradient is present.

\subsection{Data Analysis}

\begin{figure}[h!]
    \centering
    \includegraphics[width = 0.5\linewidth]{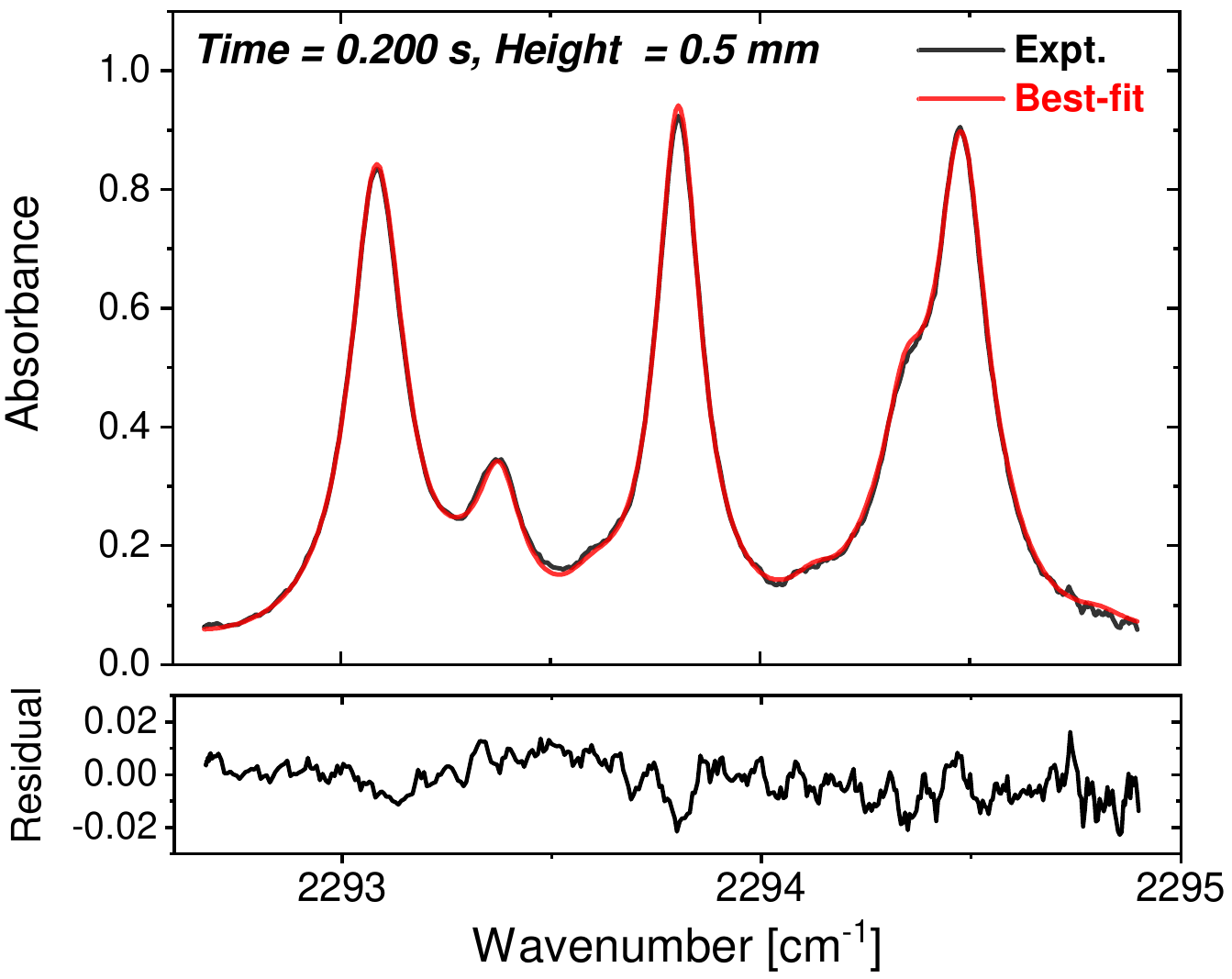}
    \caption{\footnotesize A representative CO$_2$ absorption spectrum obtained in a laser up-scan.}
    \label{Fig_03}
\end{figure}

The absorbance signal $\alpha(t)$ in each laser modulation cycle was calculated from the transmitted laser intensities $I(t)$ and $I_0(t)$ (measured with and without CO$_2$ in the beam path) using the Beer-Lambert relation, as described in Eqn. \ref{Eqn1}. This time sequence of absorbance signal was then converted to an absorbance spectrum $\alpha(\nu)$, as shown in Fig. \ref{Fig_03}, based on the wavenumber mapping $\nu(t)$ determined in the etalon measurement. 

\begin{equation} \label{Eqn1}
\begin{aligned}
\alpha(t) & = ln \big[I(t)/I_0(t)\big]
\end{aligned}
\end{equation}

From the absorbance spectrum, the CO$_2$ mole fraction, $X$, was extracted using a detailed spectroscopy model as follows:

\begin{equation} \label{Eqn2}
\begin{aligned}
\alpha(\nu) & = X P L \sum_{k = 1}^{K} S_k(T) \phi_V(\nu - \nu_{0,k}, \Delta\nu_{C,k}, \Delta\nu_{D,k}) = \alpha^0(\nu) X\\
\end{aligned}
\end{equation}

This equation has accounted for the contributions from a total number of $K = 10$ adjacent absorption transitions between 2290 and 2300 cm$^{-1}$, with $S_k$, $\nu_{0,k}$, $\Delta\nu_{D,k}$ and $\Delta\nu_{C,k}$ representing the linestrength, center wavenumber, Doppler linewidth and collisional linewidth of the $k$-th transition respectively. All of these spectroscopic parameters were obtained from the latest HITRAN database \cite{Gordon2022HITRAN2020}. These transitions were modeled with the Voigt lineshape function $\phi_V$, which was evaluated numerically using the Humlicek \cite{humlivcek1982optimized} algorithm. The absorbance $\alpha$ was proportional to the CO$_2$ mole fraction $X$. In the current study, where the temperature ($T$ = 297 K), pressure ($P$ = 1 atm), and optical path-length ($L$ = 50 mm) were fixed, the proportionality factor $\alpha^0$ was a function of wavenumber $\nu$ only. From the measured $\alpha(\nu)$, the local CO$_2$ concentration at a specific time and height can be determined by a least-square fit. The overall $1\sigma$ uncertainty in the measured concentration, calculated as the root-sum-square of the individual contributions from absorption linestrength (1\%), path length (0.5\%), total pressure (0.5\%), baseline drift (1\%), and fitting error (1\%), was less than 2\%. In addition, the nominal height of the beam center, $z$, can be determined from the galvo scanner signal $V(t)$ using the known geometric relationship of the current optical configuration.

\section{Validation Experiment}
\subsection{Results of Unsteady Gas Transport Experiment with Pure Water}
\begin{figure}[h!]
    \centering
    \includegraphics[width = \linewidth]{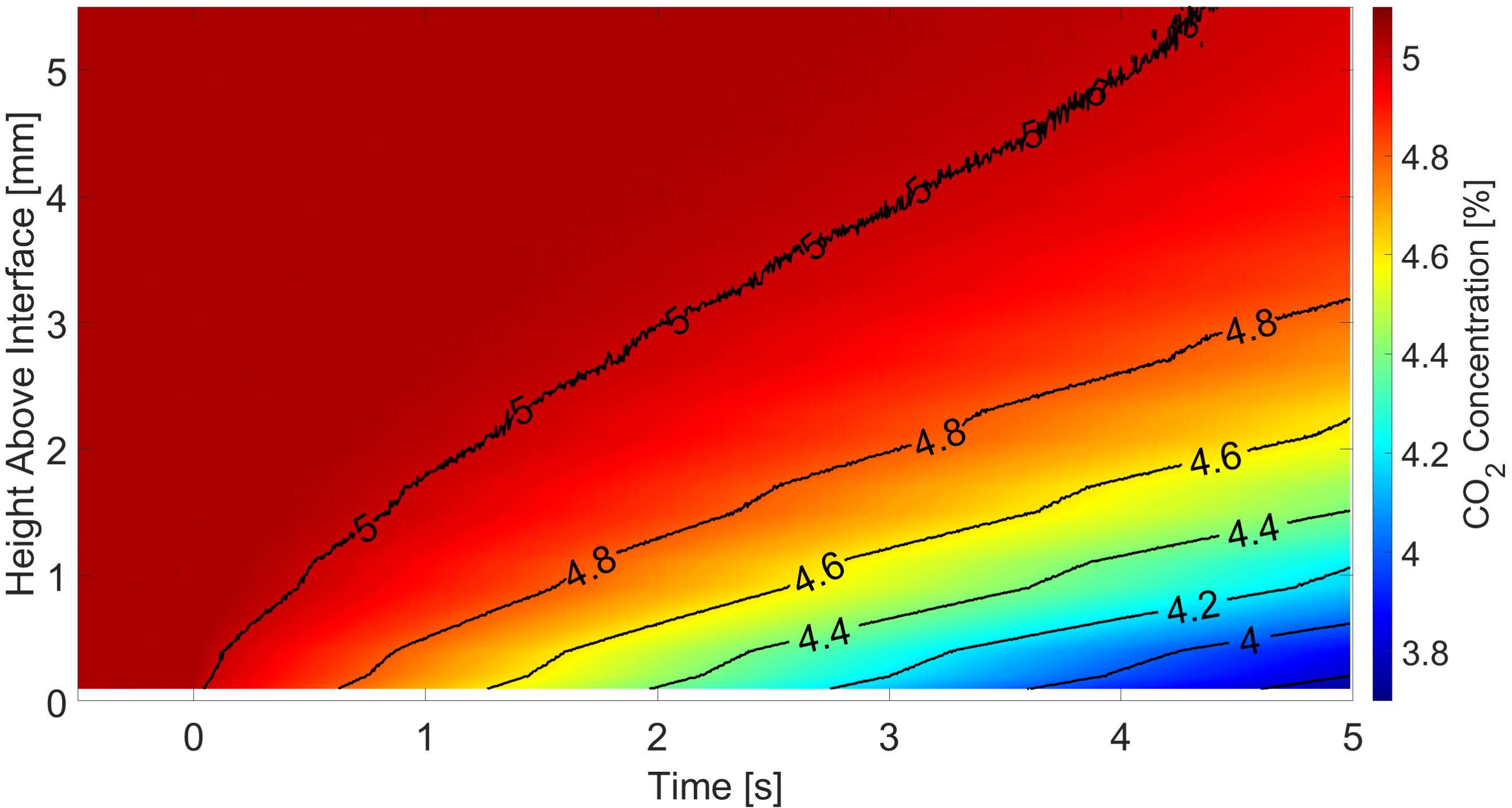}
    \caption{\footnotesize Spatiotemporal distribution of CO$_2$ mole fraction during cross-interface CO$_2$ transport experiment with pure water.}
    \label{Fig_04}
\end{figure}

\begin{figure}[h!]
    \centering
    \includegraphics[width = 0.75\linewidth]{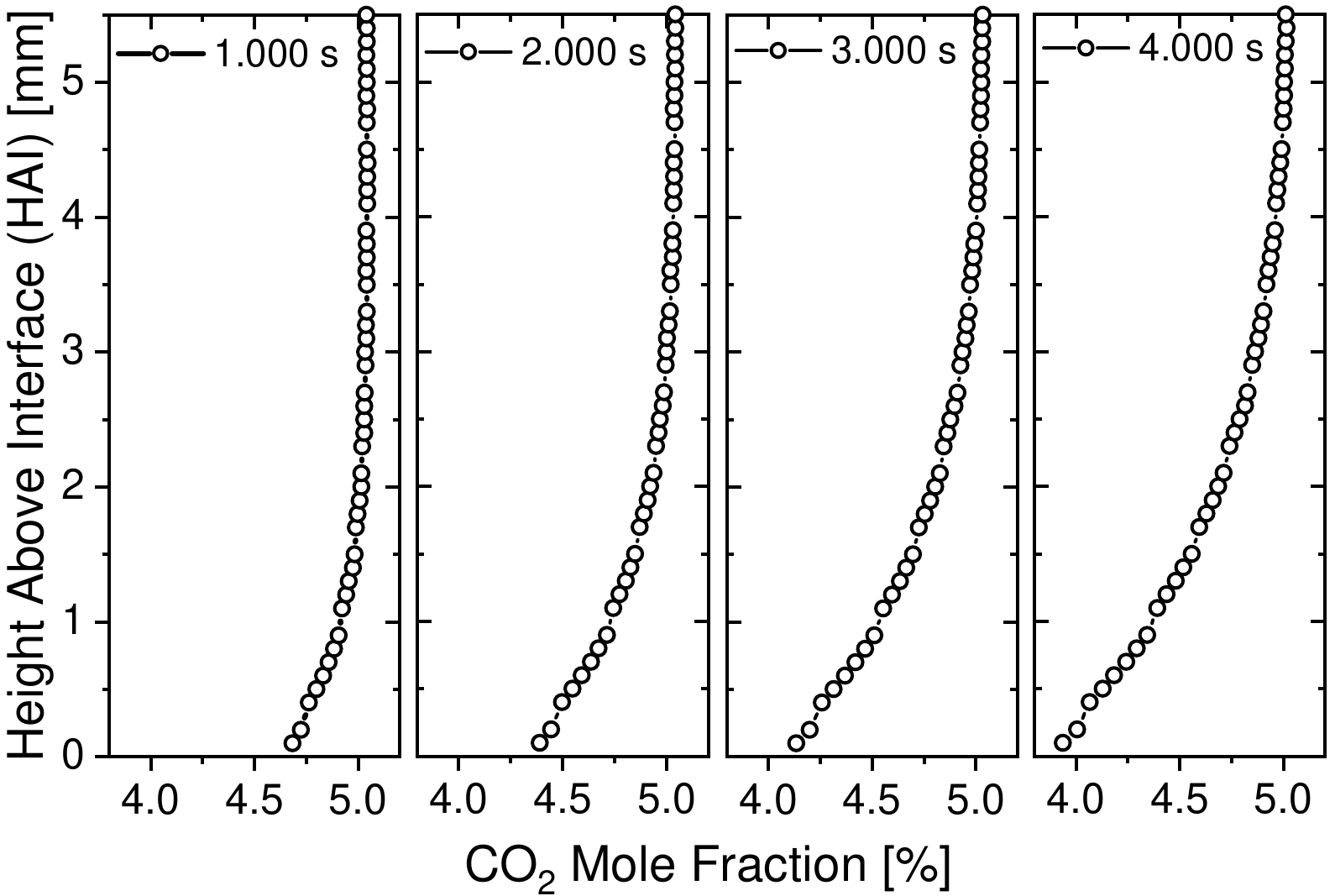}
    \caption{\footnotesize Snapshots of the CO$_2$ vertical distribution at different times.}
    \label{Fig_05}
\end{figure}

A representative set of data for the evolution of gas-phase CO$_2$ concentration distribution above the interface is illustrated in Fig. \ref{Fig_04}. Close to the interface, the CO$_2$ concentration began to decrease immediately after opening of the partition plate at time zero, as a result of absorption into the liquid phase. This concentration change created a gradient in the gas phase that drove CO$_2$ flux via molecular diffusion, and its "zone of influence" (as indicated by the constant CO$_2$ concentration contours) spread out at a rate approximately proportional to the square root of time. This can also be visulized from the vertical profiles of CO$_2$ concentration at four representative time segments (in increments of 1 second), as presented in Fig. \ref{Fig_05}.

\begin{figure}[h!]
    \centering
    \includegraphics[width = 0.5\linewidth]{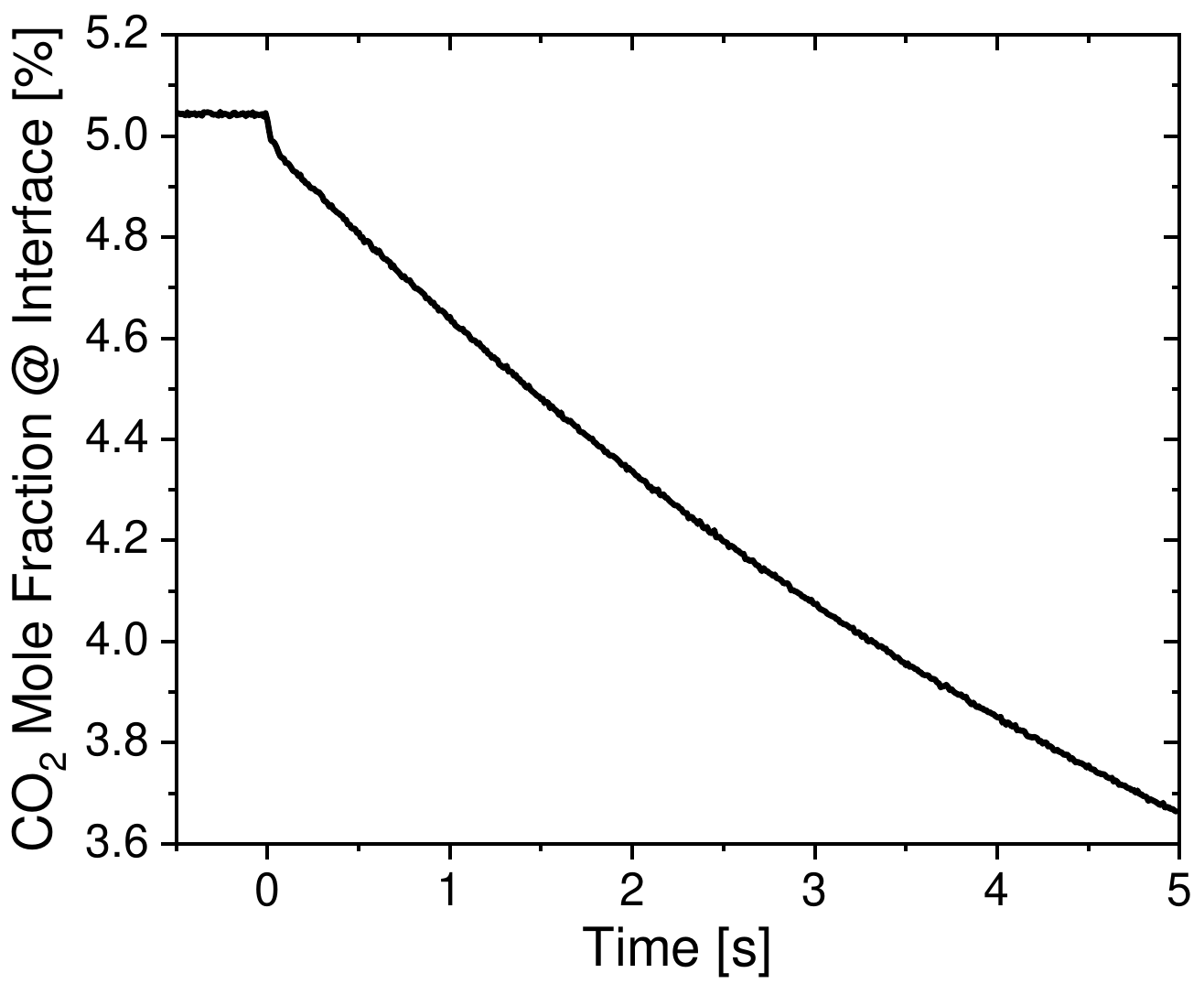}
    \caption{\footnotesize Gas-phase CO$_2$ mole fraction time-history at the interface location.}
    \label{Fig_06}
\end{figure}

Fig. \ref{Fig_06} shows the time-history of gas-phase CO$_2$ concentration at the interface, determined by extrapolating data within 0.5 mm above the interface where the laser beam was not severely blocked by water. Based on the standard deviation of the background CO$_2$ concentration before time zero, the precision of the current measurement was estimated to be 27 ppm, an order of magnitude lower than the atmospheric background carbon dioxide concentration. The interfacial CO$_2$ concentration decayed from 5.043\% to 3.660\% within 5 seconds of the experiment, which corresponded to a characteristic time of 15.60 s. This decay was resulted from the combined effects of gas diffusion and absorption, as will be further explored in the Section 4.

\subsection{Comparison with 1-D Gas Diffusion Model}

\begin{figure}[h!]
    \centering
    \includegraphics[width = \linewidth]{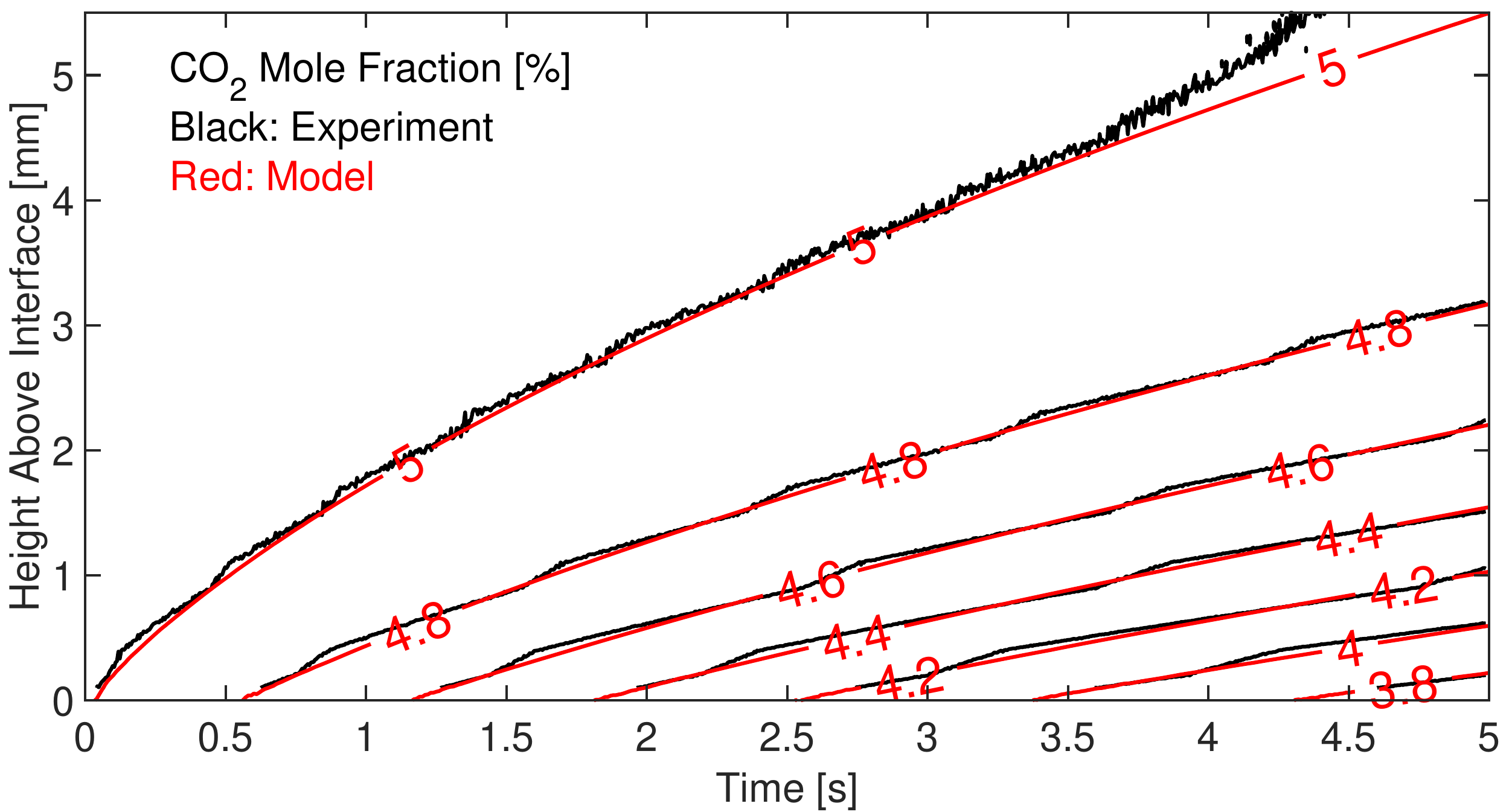}
    \caption{\footnotesize Comparison between the constant CO$_2$ concentration contours obtained from the current measurement (black) and the 1-D diffusion model (red).}
    \label{Fig_07}
\end{figure}

The current measurement results are compared with a classic 1-D gas diffusion model, as shown in Fig. \ref{Fig_07}. The basic assumptions of this model are: (1) the diffusion of CO$_2$ follows Fick's law with an effective diffusion coefficient of $D$, (2) there is no source or sink of CO$_2$ above the interface, (3) the gas density remains constant throughout the experiment, and (4) within a short test time of 5 s, the CO$_2$ concentration far away from the interface remains largely unaffected. Together, these assumptions lead to the following governing equation for the CO$_2$ mole fraction $X(z,t)$:

\begin{equation} \label{Eqn3}
\begin{aligned}
& \frac{\partial X}{\partial t} = D\frac{\partial^2X}{\partial z^2}\\
& X(0,t) = X_{0}(t)\\
& X(+\infty,t) = X(z,0) = X_{\infty}
\end{aligned}
\end{equation}

In this model, the evolution of CO$_2$ concentration distribution is driven solely by changes in the interface concentration, $X_0(t)$, whose value is set to be the same as the experimental value shown in Fig. \ref{Fig_06}. Eqn. 3 is solved numerically using an explicit finite-difference scheme, with sufficiently small spatial and temporal step sizes (on the order of 0.1 mm and 0.01 ms) to ensure numerical accuracy. Excellent agreement is observed between the current measurement results and predictions of the 1-D gas diffusion model.

\section{pH-Dependence of CO$_2$ Transport}

\begin{figure}[h!]
    \centering
    \includegraphics[width = \linewidth]{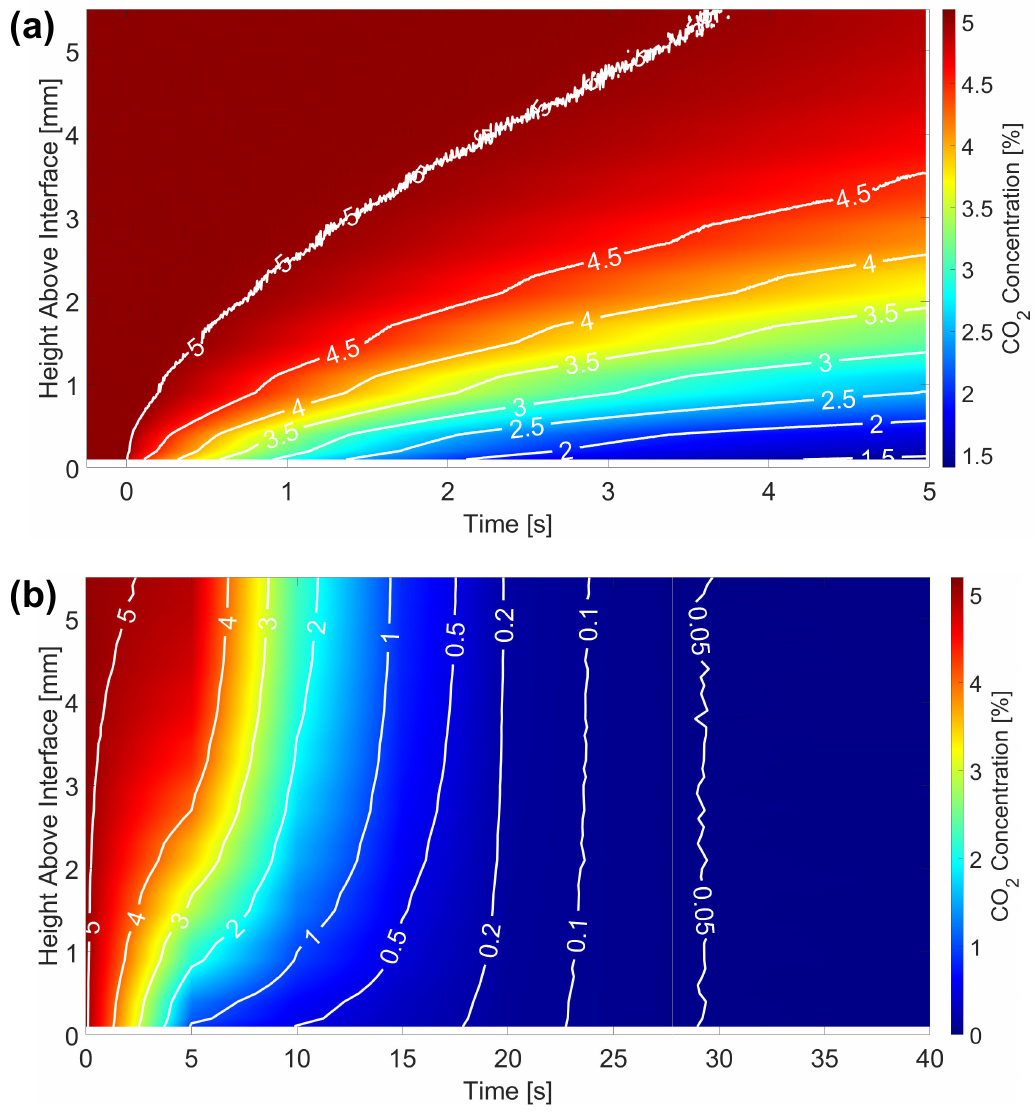}
    \caption{\footnotesize Representative measurement data of gas-phase CO$_2$ concentration evolution at pH = 9.0. (a) Short-term variation within the first 5 seconds of the experiment. (b) Long-term variation over 40 s.}
    \label{Fig_08}
\end{figure}

A series of additional experiments were also conducted for CO$_2$ cross-interface transport into alkaline solutions of pH values between 7.5 and 12.0. For the case of pH = 9.0, a set of representative data are illustrated in Fig. \ref{Fig_08}. The short-term behavior of the CO$_2$ concentration evolution appeared similar to the case of pure water, although the interface concentration decreased at a much faster rate. At longer times, due to the finite volume of the current chamber, the gas-phase CO$_2$ concentration eventually became uniform and reached an equilibrium value that was virtually zero. Results for other pH values can be found in the Supplementary Material of this paper. 

\begin{figure}[h!]
    \centering
    \includegraphics[width = 0.5\linewidth]{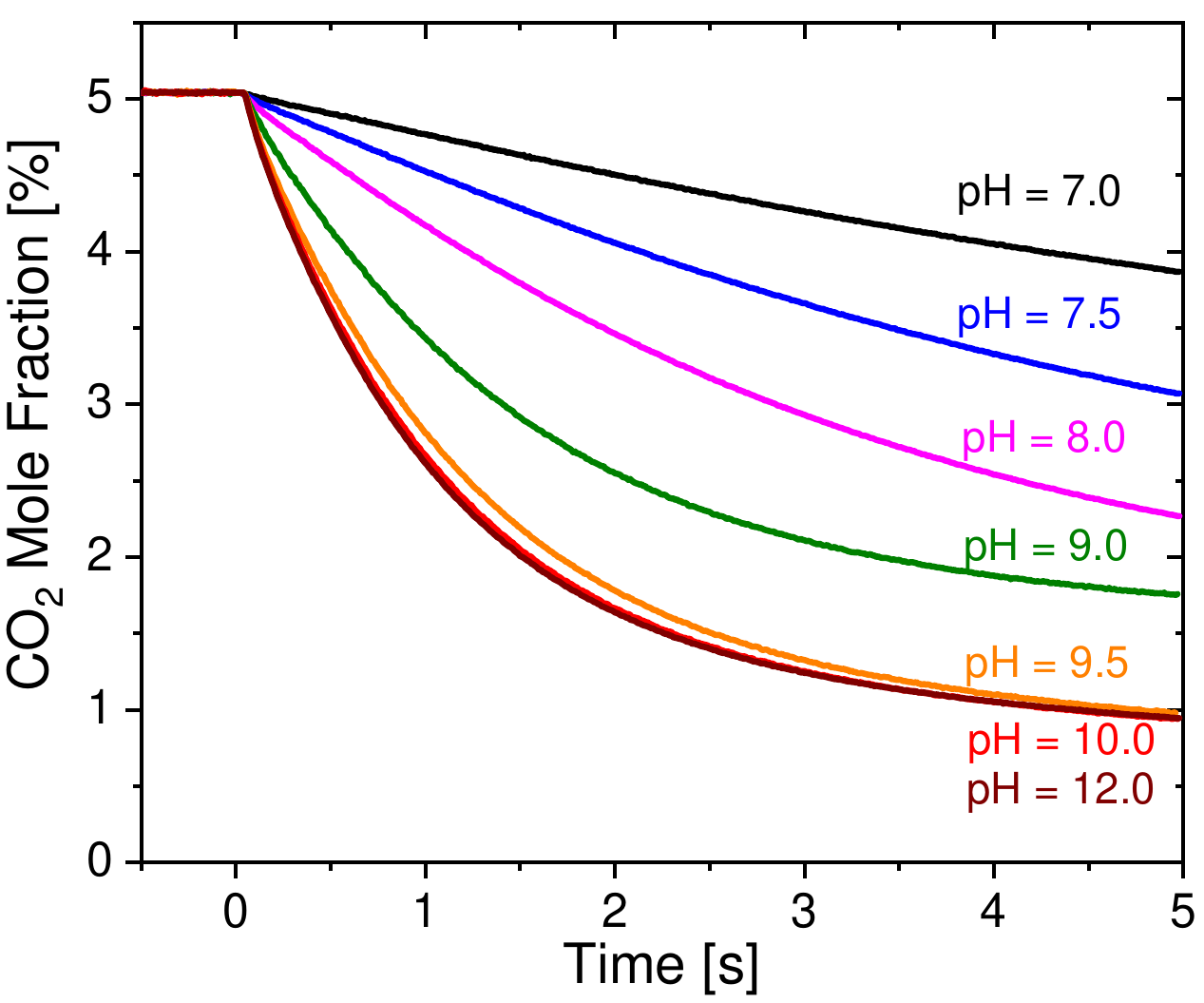}
    \caption{\footnotesize Interfacial CO$_2$ concentration at different pH.}
    \label{Fig_09}
\end{figure}

A closer examination of the depletion rate of gas-phase CO$_2$ concentration at the interface (see Fig. \ref{Fig_09}) revealed that it was strongly sensitive to pH under neutral or weakly basic conditions but quickly saturated near pH 10, reflecting competition between gas-/liquid-phase diffusion and absorption/reaction in the neighborhood of the interface. 

\begin{figure}[h!]
    \centering
    \includegraphics[width = 0.5\linewidth]{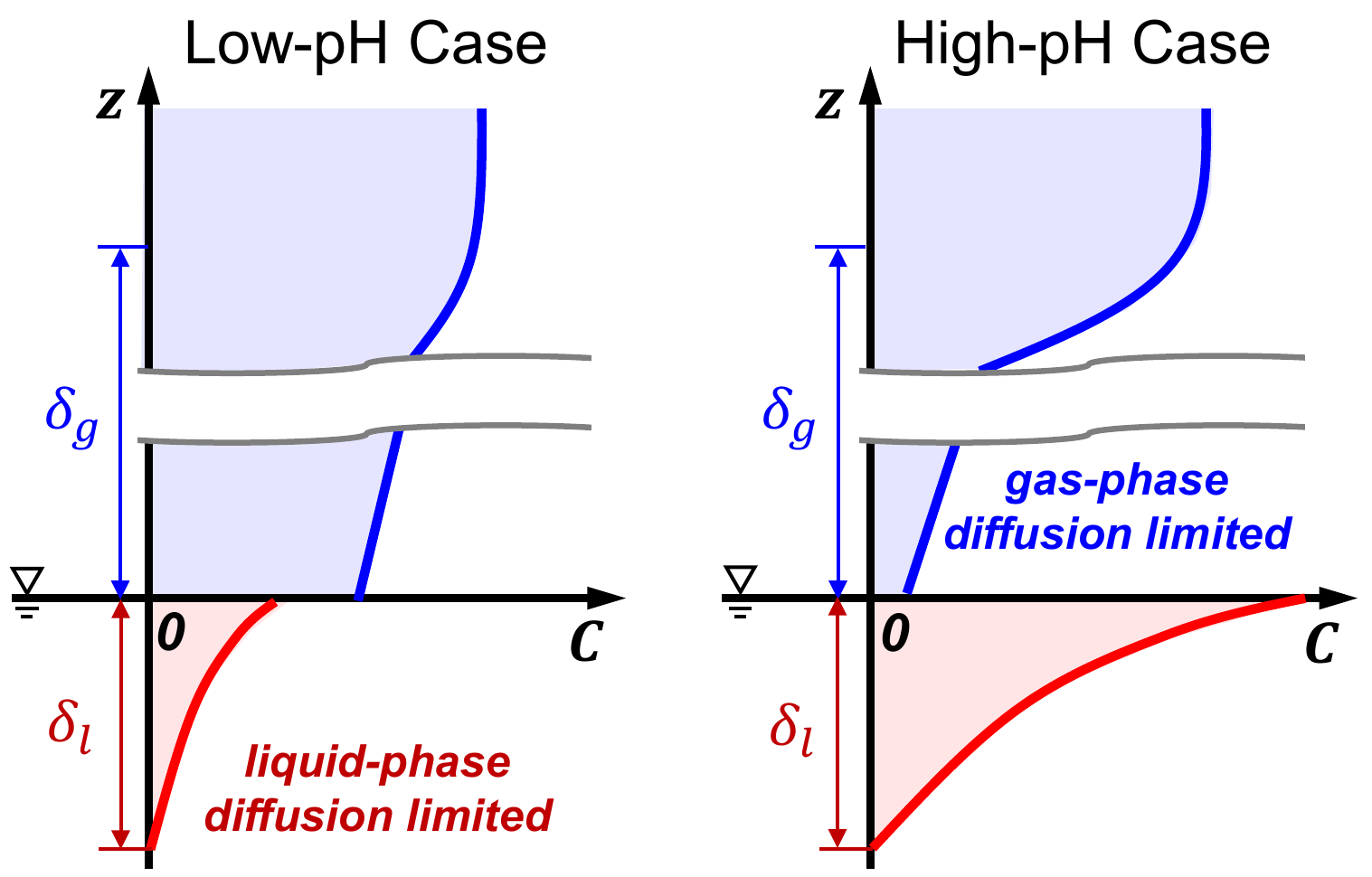}
    \caption{\footnotesize A conceptual view of the two-phase diffusion-absorption process.}
    \label{Fig_10}
\end{figure}

A possible explanation of the underlying mechanism is illustrated in Fig. \ref{Fig_10}. At low pH values, the overall rate of CO$_2$ transport is limited by diffusion in the liquid phase. Due to the relatively low solubility/saturation concentration of CO$_2$, the surface layer on the water side quickly becomes saturated soon after the beginning of interface transfer. To effectively drive the cross-interface flux, the absorbed CO$_2$ in this saturated surface layer needs to diffuse further into the bulk liquid. Because the molecular diffusivity of CO$_2$ in water is four orders of magnitude lower than that in air, diffusion in the liquid phase becomes the bottleneck.

This situation is reversed at high pH values. As the saturation concentration of CO$_2$ in the liquid phase grows exponentially with pH, the surface layer on the water side eventually becomes a virtually infinite CO$_2$ sink that instantaneously absorbs CO$_2$ upon contact, which almost completely depletes the gas-phase CO$_2$ concentration at the interface. In this scenario, the CO$_2$ flux is ultimately supplied by the bulk gas far away from the interface, and the overall rate of CO$_2$ transport is limited by gas-phase diffusion.

\begin{figure}[h!]
    \centering
    \includegraphics[width = 0.5\linewidth]{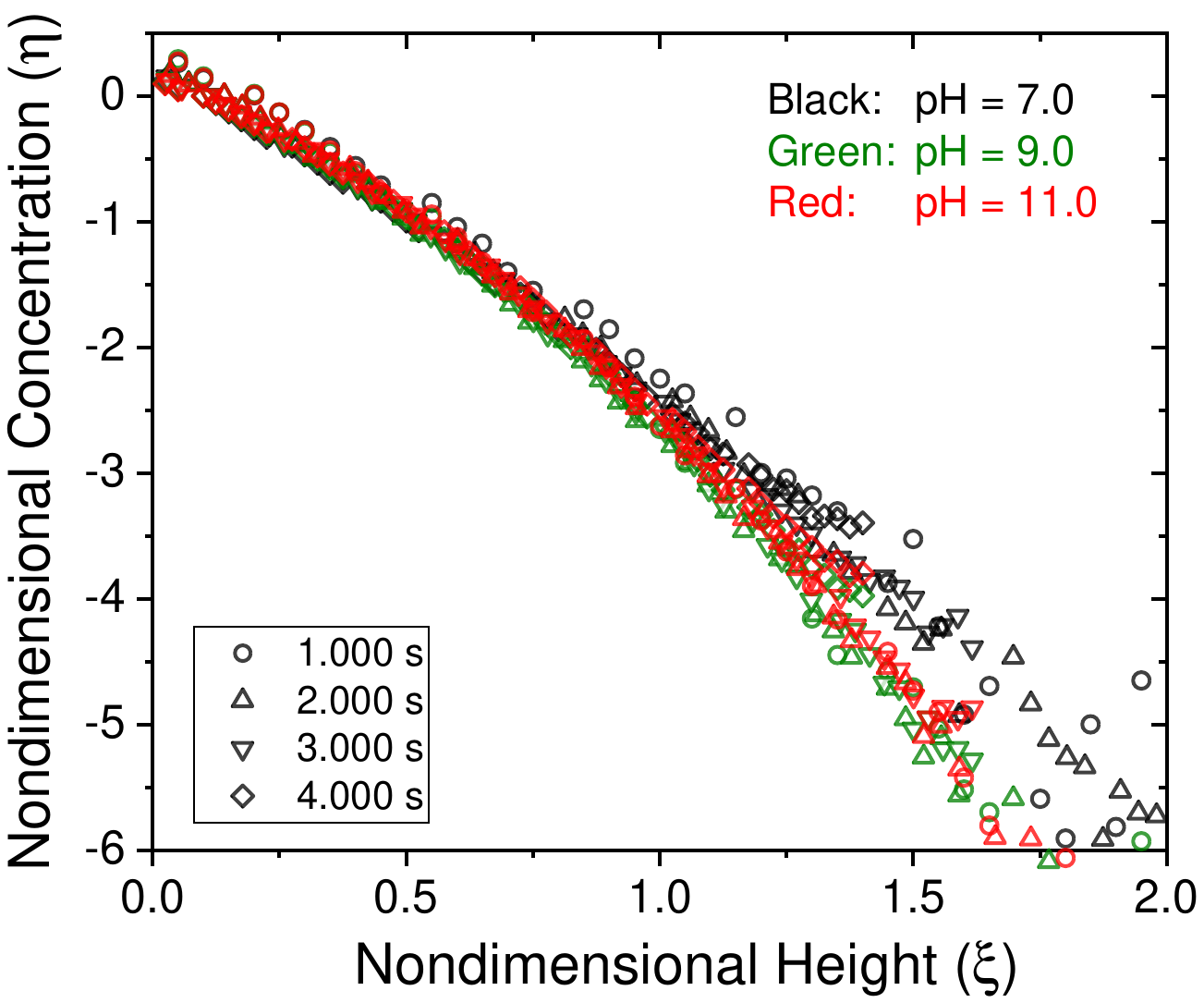}
    \caption{\footnotesize Non-dimensional distributions of CO$_2$ concentration at different pH values.}
    \label{Fig_11}
\end{figure}

It is worthy to note that, for all pH values the gas-phase CO$_2$ concentration evolution is governed by the same set of equations (i.e., Eqn. 3), which remains valid as long as the actual time-history of $X_0(t)$ is provided. In this regard, a non-dimensional analysis of the gas-phase CO$_2$ concentration distribution is performed using:

\begin{equation} \label{Eqn4}
\begin{aligned}
\xi & = \frac{z}{2D\sqrt{t}} \\
\eta & = ln\bigg[ \frac{X_\infty-X(z,t)}{X_\infty - X_0(t)} \bigg]
\end{aligned}
\end{equation}

where $\xi$ and $\eta$ represent the non-dimensional height and concentration, respectively. Fig. \ref{Fig_11} shows the non-dimensional concentration profiles at four representative times (1.000 s, 2.000 s, 3.000 s, and 4.000 s) and three pH values (7.0, 9.0 and 11.0). All data appear to collapse onto a single curve, suggesting that a universal scaling law exists for the evolution of gas-phase CO$_2$ distribution. This topic may merit further investigation in future studies, perhaps under more complex conditions of gas transfer and in conjunction with liquid-phase diagnostics as well.

\section{Conclusions}

This study established an in situ method for quantitative measurement of carbon dioxide concentration distribution at the air-water interface, based on tunable laser absorption spectroscopy and rapid beam scanning. Validation experiments were conducted in a controlled laboratory environment of a quasi-1D gas diffusion chamber, where the diffusion and absorption of gas-phase CO$_2$ across the interface into pure water were continuously monitored, with an effective time resolution of 5 ms and an effective spatial resolution of 1 mm, respectively. The spatiotemporal evolution of gas-phase CO$_2$ concentration was seen in good agreement with the classic 1-D diffusion model. The precision of CO$_2$ measurement was approximately 27 ppm, while the overall uncertainty of the measurement was less than $2\%$.

Additional experiments were conducted with alkaline solutions of pH values between 7.5 and 12.0. The depletion rate of gas-phase CO$_2$ concentration at the interface exhibited strong pH sensitivity at relatively low pH but saturated near pH 10, revealing a complex competition between gas-/liquid-phase diffusion and absorption/reaction occurring in the immediate vicinity of the interface. At high pH values, the effective solubility of CO$_2$ became sufficiently large so that the dominant resistance to mass transfer shifted to the gas phase. In this sense, these experiments have also provided a platform to study the effect of solubility on gas transfer rates.

Because of its high temporal and spatial resolutions, the current method promises strong utility for experimental investigations of cross-interface gas transport phenomena, including both field measurements and laboratory studies under more complex flow conditions. Potential applications include fundamental air–sea gas transport studies in physical oceanography and practical implementations of advanced CO$_2$ monitoring in modern carbon sequestration technologies.

\begin{acknowledgments}
This research is supported by the Marine S\&T Fund of Shandong Province for Laoshan Laboratory under Grant No. 2022LSL010201-3 and the National Natural Science Foundation of China under Grants No. 12472278 and No. 92152108.
\end{acknowledgments}

\bibliography{References}
\end{document}